\documentclass[aps,prl,twocolumn,groupedaddress,showpacs,showkeys]{revtex4}

\usepackage{graphicx}
\usepackage{bm}
\usepackage{amsmath}

\begin{document}

\title{Nonlinear Self-Trapping of Matter Waves in Periodic Potentials}

\author{Th. Anker$^1$, M. Albiez$^1$, R. Gati$^1$, S. Hunsmann$^1$,  B. Eiermann$^1$, A. Trombettoni$^2$ and M.K. Oberthaler$^1$}

\affiliation{$^1$ Kirchhoff Institut f\"ur Physik, Universit{\"a}t
Heidelberg, Im Neuenheimer Feld 227, 69120 Heidelberg, Germany \\
$^2$ I.N.F.M. and Dipartimento di Fisica, Universit\'a di Parma,
parco Area delle Scienze 7A, I-43100 Parma, Italy}

\date{\today}

\begin{abstract}
We report the first experimental observation of nonlinear
self-trapping of Bose-condensed $^{87}$Rb atoms in a one dimensional
waveguide with a superimposed deep periodic potential . The trapping
effect is confirmed directly by imaging the atomic spatial
distribution. Increasing the nonlinearity we move the system from
the diffusive regime, characterized by an expansion of the
condensate, to the nonlinearity dominated self-trapping regime,
where the initial expansion stops and the width remains finite. The
data are in quantitative agreement with the solutions of the
corresponding discrete nonlinear equation. Our results reveal that
the effect of nonlinear self-trapping is of local nature, and is
closely related to the macroscopic self-trapping phenomenon already
predicted for double-well systems.
\end{abstract}


\maketitle

The understanding of coherent transport of waves is essential for
many different fields in physics. In contrast to the dynamics of
non-interacting waves, which is conceptually simple, the situation
can become extremely complex as soon as interaction between the
waves is of relevance. Very intriguing and counter intuitive
transport phenomena arise in the presence of a periodic potential.
This is mainly due to the existence of spatially localized
stationary solutions.

In the following we will investigate the dynamics of Bose-condensed
$^{87}$Rb atoms in a deep one dimensional periodic potential, i.e.
the matter waves are spatially localized in each potential minimum
(tight binding) and are coupled via tunneling to their next
neighbors. This system is described as an array of coupled Boson
Josephson junctions \cite{cataliotti:2001:josephson_junctions}. The
presence of nonlinearity drastically changes the tunneling dynamics
\cite{raghavan_1999_prlpra} leading to new localization phenomena on
a macroscopic scale such as discrete solitons, i.e. coherent
non-spreading wave packets, and nonlinear self-trapping
\cite{Trombettoni01}. These phenomena have also been studied in the
field of nonlinear photon optics where a periodic refractive index
structure leads to an array of wave guides, which are coupled via
evanescent waves \cite{christodoulides:2003:nonlinearwaveguides}.

\begin{figure}[ht!]
\includegraphics[width=7.3cm]{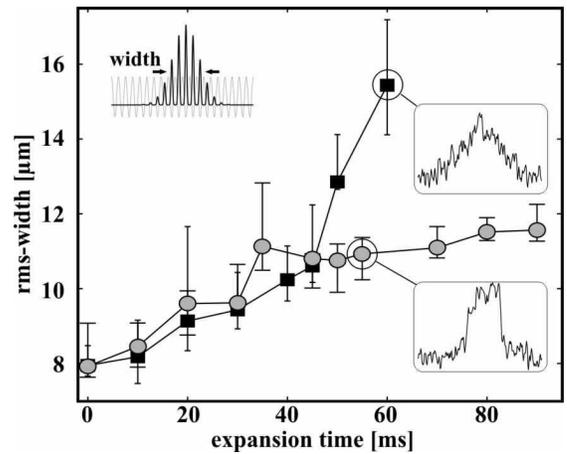}
\caption{\label{fig:1} Observation of nonlinear self-trapping of
Bose-condensed $^{87}$Rb atoms. The dynamics of the wave packet
width along the periodic potential is shown for two different
initial atom numbers. By increasing the number of atoms from
2000$\pm$200(squares) to 5000$\pm$600(circles), the repulsive
atom-atom interaction leads to the stopping of the global expansion
of the wave packet. The insets show that the wave packet remains
almost gaussian in the diffusive regime but develops steep edges in
the self-trapping regime. These edges act as boundaries for the
complex dynamics inside.}
\end{figure}

In this letter we report on the first experimental confirmation of
the theoretically predicted effect of nonlinear self-trapping of
matter waves in a periodic potential \cite{Trombettoni01}. This
effect describes the drastic change of the dynamics of an expanding
wave packet, when the nonlinearity i.e. repulsive interaction
energy, is increased above a critical value. Here the
counterintuitive situation arises that although the spreading is
expected to become faster due to the higher nonlinear pressure, the
wave packet {\em stops} to expand after a short initial diffusive
expansion. Since we observe the dynamics in real space, we can
directly measure the wave packet width for different propagation
times. In Fig.~\ref{fig:1} we show the experimental signature of the
transition from the diffusive to the self-trapping regime. We
prepare wave packets in a periodic potential and change only the
nonlinear energy by adjusting the number of atoms in the wave packet
close to (2000$\pm$200 atoms) and above (5000$\pm$600 atoms) the
critical value. Clearly both wave packets expand initially. At
$t\sim$35 {\it m}s the wave packet with higher initial atomic
density has developed steep edges and stops expanding (see inset in
Fig.~\ref{fig:1}). In contrast, the wave packet with the lower
initial atomic density continues to expand keeping its gaussian
shape.

The coherent matter-wave packets are generated with $^{87}$Rb
Bose-Einstein condensates realized in a crossed light beam dipole
trap ($\lambda$ = 1064{\it n}m, 1/$e^2$ waist 55$\mu$m, 600{\it
m}W per beam). Subsequently a periodic dipole potential $V_p =
s\cdot E_{r}\sin^2(k x)$, realized with a far off-resonant
standing light wave ($\lambda = 783${\it n}m) collinear with one
of the dipole trap beams is adiabatically ramped up . The depth of
the potential is proportional to the intensity of the light wave
and is given in recoil energies $E_r=\frac{\hbar^2k^2}{2m}$ with
the wave vector $k=2\pi/\lambda$. By switching off the dipole trap
beam perpendicular to the periodic potential the atomic matter
wave is released into a trap acting as a one-dimensional waveguide
$V_{dip}=\frac{m}{2}(\omega_\perp^2r^2 + \omega_\parallel^2x^2)$
with radial trapping frequency $\omega_\perp = 2 \pi \cdot 230
\,\text{Hz}$ and longitudinal trapping frequency $\omega_\parallel
\approx 2 \pi \cdot 1 \,\text{Hz}$. The wave packet evolution
inside the combined potential of the waveguide and the lattice is
studied by taking absorption images of the atomic density
distribution after a variable time delay. The density profiles
$n(x,t)$ along the waveguide are obtained by integrating the
absorption images over the radial dimensions and allow the
detailed investigation of the wave packet shape dynamics with a
spatial resolution of 3$\mu$m.

\begin{figure}[h]
\includegraphics[width=8.5cm]{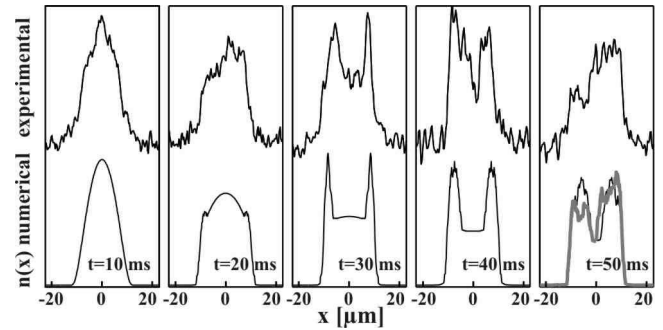}
\caption{\label{fig:3} Comparison between theory and experiment for
$s$ = 10, 7.6(5) $\mu$m initial rms-width, and 5000$\pm$600 atoms.
The upper graphs show the measured density distribution for
different propagation times. During the initial expansion in the
self-trapping regime the wave packet develops steep edges which act
as stationary boundaries for the subsequent internal dynamics. The
results of the numerical integration of eq.~\ref{DNL} (depicted in
the lower graphs) are in very good agreement. For $t = 50$ {\it m}s
a 1.5 {\it m}rad deviation of the wave guides's horizontal
orientation (consistent with the experimental uncertainty) is taken
into account and reproduces the experimentally observed asymmetry
(gray line).}
\end{figure}

In Fig.~\ref{fig:3} the measured temporal evolution of the wave
packet prepared in the self-trapping regime ($s$ = 10, 7.6(5) $\mu$m
initial rms-width, 5000$\pm$600 atoms) is shown. The evolution of
the shape is divided into two characteristic time intervals.
Initially ($t < 20$ {\it m}s) the wave packet expands and develops
steep edges. This dynamics can be understood in a simple way by
considering that the repulsive interaction leads to a broadening of
the momentum distribution and thus to a spreading in real space.
Since the matter waves propagate in a periodic potential the
evolution is governed by the modified dispersion (i.e. band
structure) $E(q) = -2K\cos(dq)$ where $d = \lambda/2$ is the lattice
spacing, $\hbar q$ is the quasimomentum and $K$ is the
characteristic energy associated with the tunneling. The formation
of steep edges is a consequence of the population of higher
quasimomenta around $q =\pm \pi/2d$ where the dispersion is strongly
reduced and the group velocity is extremal. In order to populate
quasimomenta  $|q| > \pi/2d$ the initial interaction energy has to
be higher than the characteristic tunneling energy $K$ and thus the
critical parameter depends on the ratio between the on-site
interaction energy and the tunneling energy as we will discuss in
detail. While in the linear evolution the steep edges move with the
extremal group velocity \cite{dispersionmanagement}, in the
experiment reported here they stop after their formation. As we will
show this is a consequence of the high atomic density gradient at
the edge which suppresses tunneling between neighboring wells. The
further evolution is characterized by stationary edges acting as
boundaries for the complex internal behavior of the wave packet
shape. The formation of the side peaks is an indication that atoms
moving outwards are piled up because they cannot pass the steep
edge. Finally the pronounced features of the wave packet shape
disappear and a square shaped density distribution is formed.

In order to understand in detail the ongoing complex self-trapping
dynamics we compare quantitatively our experimental findings with
numerically obtained solutions (see Fig. \ref{fig:3}).  For our
typical experimental parameters of $s\sim11$ and $\sim100$ atoms per
well we are in the regime where the dynamics can be described by a
macroscopic wave function $\Psi(\vec{r},t)$ and thus by the
Gross-Pitaevski equation (GPE) \cite{zwerger_03_mott_hubbard}. Since
we use deep optical lattices the description can be reduced to a one
dimensional discrete nonlinear equation, which includes the
fundamental processes, namely tunneling between the wells and
nonlinear phase evolution due to the interaction of the atoms
\cite{Trombettoni01,DNL}. In our experiment the trapping frequency
in a single well along the lattice period is on the order of
$\omega_x\approx 2 \pi \cdot 25$ {\it k}Hz, whereas the transverse
trapping frequency of the wave guide is $\omega_\perp = 2 \pi \cdot
230 \,\text{Hz}$. Thus our system can be described as a horizontal
pile of pancakes, and the transverse degree of freedom cannot be
neglected. In \cite{DNL} a one dimensional discrete nonlinear
equation (DNL) is derived which takes into account the adiabatic
change of the wave function in the transverse direction due to the
atom-atom interaction. A generalized tight binding ansatz
\begin{equation}
\Psi(\vec{r},t) = \sum_j \psi_j(t)\Phi_j(\vec{r},N_j(t))\label{tba}
\end{equation}
is used, with $\psi_j(t) = \sqrt{N_j(t)}e^{i\phi_j(t)}$, where
$N_j(t)$ is the atom number and $\phi_j(t)$ is the phase of the
$j$th condensate.  $\Phi_j$ is normalized to 1 (i.e. $\int d\vec{r}
\Phi_j^2=1$) and $\Psi(\vec{r},t)$ is normalized to the total number
of atoms $N_T$ (i.e. $\sum_j | \psi_j |^2 = N_T$). The spatial real
wave function $\Phi_j(\vec{r},N_j(t))$ is centered at the minimum of
the j-th well and is time dependent through $N(t)$. Integrating over
the spatial degrees of freedom, the following DNL is obtained from
the GPE :
\begin{eqnarray}\label{DNL}
i\hbar\frac{\partial\psi_j}{\partial t}&=&
\epsilon_j\psi_j-K(\psi_{j+1} + \psi_{j-1}) + \mu_j^{loc}\psi_j .
\end{eqnarray}
$K$ is the characteristic tunneling energy between adjacent sites.
$\epsilon_j=\int d\vec{r}\frac{m}{2}\omega^2_\parallel x^2 \Phi^2_j$
is the on-site energy resulting from the longitudinal trapping
potential, which is negligible in the description of our experiment.
The relevant chemical potential is given by $\mu_j^{loc} = \int
d\vec{r} \left[ \frac{m}{2}\omega_\perp^2 r^2\Phi^2_j + g_0
|\psi_j(t)|^2\Phi^4_j \right]$ with $g_0=4\pi\hbar^2a/m$ ($a$ is the
scattering length). It can be calculated approximately for our
experimental situation assuming a parabolic shape in transverse
direction (Thomas-Fermi approximation) and a Gaussian shape in
longitudinal direction for $\Phi_j(\vec{r},N_j(t))$ ($\omega_{x}\gg
\mu_j^{loc}/\hbar
>\omega_\perp$). This leads to $\mu_j^{loc} = U_1 |\psi_j(t)|$ with
\begin{equation}
U_1 = \sqrt{\frac{m\omega_\perp^2
g_0}{\sqrt{2\pi}\pi\sigma_x}}\label{U1}.
\end{equation}
Here $\sigma_x = \lambda/(2\pi s^{\frac{1}{4}})$ is the
longitudinal Gaussian width of $\Phi_j$ in harmonic approximation
of the periodic potential minima. Please note, that if the local
wave function $\Phi_j$ does not dependent on $N_j$ eq.~\ref{DNL}
reduces to the well known discrete nonlinear Schr\"odinger
equation with $\mu_j^{loc} \propto N_j$
\cite{hennig99,Trombettoni01}.

We compare the experimental and numerical results in
Fig.~\ref{fig:3} and find very good agreement. The theory reproduces
the observed features such as steepening of the edges, the formation
of the side peaks and the final square wave packet shape. It is
important to note that all parameters entering the theory (initial
width, atom number, periodic potential depth and transverse trapping
frequency) have been measured independently. The observed asymmetry
of the wave packet shapes (e.g. see Fig.~\ref{fig:3}, $t$ = 50{\it
m}s) appears due to the deviation from the perfect horizontal
orientation of the wave guide ($\pm 2${\it m}rad) which results from
small changes in height of the pneumatic isolators of the optical
table during the measurements.
\begin{figure}[h]
\includegraphics[width=8.1cm]{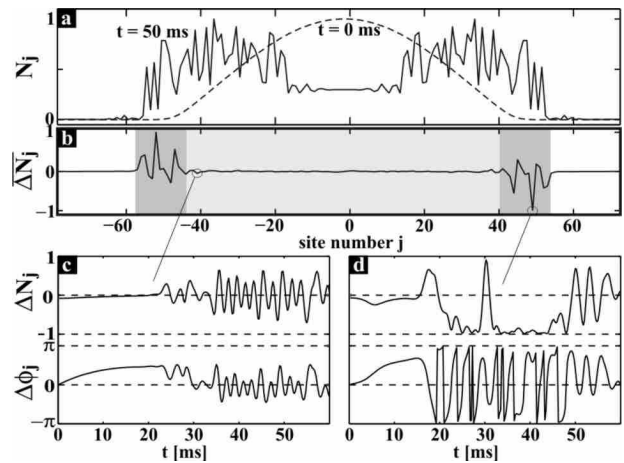}
\caption{\label{fig:4}  A numerical investigation of the site to
site tunneling dynamics. (a) The atomic distribution $N_j$ of the
wave packet for $t = 0$ and 50 {\it m}s. (b) The relative population
difference $\Delta N_j$ time averaged over the expansion time
indicates two regions with different dynamics. (c) The dynamics of
$\Delta N_j$ and the phase difference $\Delta \phi_j$ for the marked
site oscillate around zero known as the zero-phase mode of the Boson
Josephson junction. (d) The dynamics in the edge region is
characterized by long time periods where $|\Delta N_j|$ is close to
1 while at the same time $\Delta \phi_j$ winds up very quickly
(phase is plotted modulo $\pi$) known as 'running phase
self-trapping mode' in Boson Josephson junctions. Thus the expansion
of the wave packet is stopped due to the inhibited site to site
tunneling at the edge of the wave packet.}
\end{figure}

In the following we will use the numerical results to get further
insight into the self-trapping dynamics. We investigate the local
tunneling dynamics and phase evolution by evaluating the relative
atom number difference $\Delta N_j = (N_{j+1}-N_j)/(N_{j+1}+N_j)$
and the phase difference $\Delta \phi_j= \phi_{j+1} - \phi_j$
between two neighboring sites. In Fig.~\ref{fig:4}a) the wave packet
shapes for $t = 0$ and $t = 50${\it m}s are shown. In
Fig.~\ref{fig:4}b) we plot the relative atom number difference
$\Delta N_j$ averaged over the whole propagation duration of 50{\it
m}s. The graph indicates two spatial regions with different
characteristic dynamics. While the average vanishes in the central
region (shaded in light gray) it has significant amplitude in the
edge region (shaded in dark gray). The characteristic dynamics of
$\Delta N_j$ and $\Delta \phi_j$ in the central region is depicted
in Fig.~\ref{fig:4}c). The atom number difference as well as the
phase difference oscillate around zero. This behavior is known in
the context of BEC in double-well potentials. It is described as the
Boson Josephson junction 'zero-phase mode'
\cite{raghavan_1999_prlpra} and is characteristic for superfluid
tunneling dynamics if the atom number difference stays below a
critical value. At the edge in contrast, $\Delta N_j$ crosses the
critical value during the initial expansion (steep density edge) and
locks for long time periods to high absolute values showing that the
tunneling and thus the transport is inhibited. At the same time the
phase difference winds up. This characteristic dynamics has been
predicted within the Boson Josephson junction model for a
double-well system and is referred to as the 'running phase
self-trapping mode' \cite{raghavan_1999_prlpra}. This analysis makes
clear that the effect of nonlinear self-trapping as observed in our
experiment is a {\em local} effect and is closely related to Boson
Josephson junctions dynamics in a double-well system.

\begin{figure}[h]
\includegraphics[width=7.2cm]{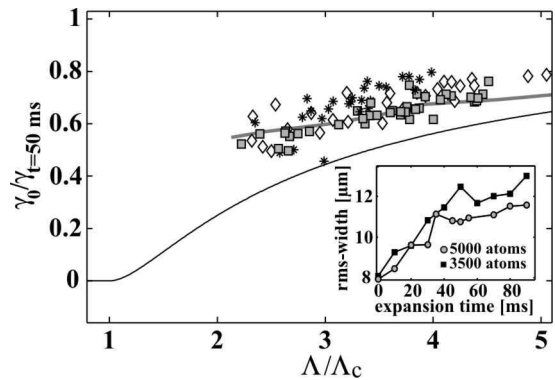}
\caption{\label{fig:5} Experimental investigation of the scaling
behavior. The solid line shows the curve given by eq.~\ref{scaling}.
Experimentally the parameter $\Lambda/\Lambda_c$ was varied by using
three different periodic potential depths: $s$ = 10.6(3) (stars),
11.1(3) (squares) and 11.5(3) (diamonds). For each potential depth
wave packets with different atom numbers and initial widths are
prepared and the width for $t = 50$ {\it m}s is determined. The
experimental data show qualitatively the scaling behavior predicted
by eq.~\ref{scaling} and are in quantitative agreement with the
results of the numerical integration of the DNL (gray line). The
inset depicts the nature of the scaling: increasing
$\Lambda/\Lambda_c$ (by e.g. increasing the atom number) leads to a
faster trapping and thus to a smaller final width.}
\end{figure}

Although the local dynamics just described is very complex, the
evolution of the root mean square width of the wave packet, i.e. the
global dynamics, can be predicted analytically within a very simple
model. In \cite{Trombettoni01} a Gaussian profile wave packet
$\psi_j(t)\propto
\exp(-(\frac{j}{\gamma(t)})^2+i\frac{\delta(t)}{2}j^2)$
parameterized by the width $\gamma(t)$ (in lattice units) and the
quadratic spatial phase $\delta(t)$, is used as an ansatz for
quasimomentum $q=0$ to solve the discrete nonlinear Schr\"odinger
equation. The time evolution of the width $\gamma(t)$ is obtained
analytically applying a variational principle. The result of this
simple model is, that the dynamics of the wave packet width is
solely determined by two global parameters - the density of the
atoms and the depth of the periodic potential. Also a critical
parameter $\Lambda/\Lambda_c$ can be deduced, which governs the
transition from the diffusive to the self-trapping regime. The
transition parameter $\Lambda/\Lambda_c$ for the 2D case described
by eq.~\ref{DNL} is obtained following the same lines of calculation
as in \cite{Trombettoni01}. Assuming that the initial width
$\gamma_0 \gg 1$ (in the experiment typically $\gamma_0\approx$40)
we obtain
\begin{eqnarray*}
\Lambda  =  \frac{U_1\sqrt{N_T}}{2K} \hspace{0.3333cm}
\text{and}\hspace{0.3333cm} \Lambda_c =
\frac{3}{2}\left(\frac{9\pi}{8}\right)^{\frac{1}{4}}\sqrt{\gamma_0}.
\end{eqnarray*}
A surprising result of this model is the prediction of the
following scaling behavior (shown in Fig.~\ref{fig:5}):
\begin{equation}
\frac{\gamma_0}{\gamma_{\infty}}= \sqrt{1-\frac{\Lambda_c}{\Lambda}}
\label{scaling}
\end{equation}
for $\Lambda/\Lambda_c>1$, where $\gamma_{\infty}$ is the width of
the wave packet for $t\rightarrow \infty$. For $\Lambda/\Lambda_c<1$
the width is not bound and thus the system is in the diffusive
regime. In the regime $\Lambda/\Lambda_c>1$ the width is constant
after an initial expansion (see inset Fig.~\ref{fig:5}). Since
$\Lambda/\Lambda_c\propto\mu_{av}^{loc}/K$, the self-trapping regime
is reached by either reducing the initial width, increasing the
height of the periodic potential or, as is shown in
Fig.~\ref{fig:1}, by increasing the number of atoms.

Scaling means that all data points (i.e. different experimental
settings with the same $\Lambda/\Lambda_c$) collapse onto a single
universal curve. In order to confirm the scaling behavior
experimentally we measure the width of the wave packet after
50{\it m}s evolution for different system parameters, i.e. atom
number, initial width of the wave packet, and depth of the
periodic potential. The experimental results shown in
Fig.~\ref{fig:5} confirm the universal scaling dependence on
$\Lambda/\Lambda_c$ and follow qualitatively the prediction of the
simple model. The dashed line in Fig.~\ref{fig:5} is the result of
the numerical integration of the discrete nonlinear equation given
in eq.~\ref{DNL} evaluated at $t=$50{\it m}s. It shows
quantitative agreement with the experiment. The difference between
the numerical (gray line) and analytical calculation (solid line)
is due to the initial non-gaussian shape (numerically obtained
ground state) and the strong deviation from the gaussian shape for
long propagation times.

Concluding we have demonstrated for the first time the predicted
effect of nonlinear self trapping of Bose-Einstein condensates in
deep periodic potentials. The detailed analysis shows that this is a
{\em local} effect, which occurs due to nonlinearity induced
inhibition of site to site tunneling at the edge of the wave packet.
This behavior is closely connected to the phenomenon of macroscopic
self trapping known in the context of double-well systems.
Furthermore we quantitatively confirm in our experiments the
predicted critical parameter which discriminates between diffusive
and self trapping behavior.

We wish to thank A. Smerzi for very stimulating discussions. This
work was supported by the Deutsche Forschungsgemeinschaft, Emmy
Noether Programm, and by the European Union, RTN-Cold Quantum
Gases, Contract No. HPRN-CT-2000-00125.

\end{document}